\documentclass[submission,copyright,creativecommons]{eptcs}
\providecommand{\event}{41st International Conference on Logic
  Programming (ICLP 2025)} 

\usepackage{iftex}

\ifpdf
  \usepackage{underscore}         
  \usepackage[T1]{fontenc}        
\else
  \usepackage{breakurl}           
\fi

\usepackage{xspace}
\usepackage{color}
\usepackage{enumitem}
\usepackage{amsmath}
\usepackage{amssymb}
\usepackage{multirow}
\usepackage{booktabs}
\usepackage{xspace}
\usepackage{tikz}
\usepackage{caption}
\usepackage{algorithm}
\usepackage[noend]{algpseudocode}
\usetikzlibrary{arrows.meta, positioning, calc,shapes.geometric,automata,patterns}
\usepackage{enumitem}
\usepackage{pgfplots}
\usepackage{subcaption}
\usepackage{wrapfig}
\usepackage{xcolor}
\usepackage{verbatim}

\newtheorem{definition}{Definition}
\newtheorem{example}{Example}
\definecolor{shadecolor}{gray}{1.00}
\definecolor{darkgray}{gray}{0.30}
\definecolor{violet}{rgb}{0.56, 0.0, 1.0}
\definecolor{forestgreen}{rgb}{0.13, 0.55, 0.13}
\definecolor{mygray}{rgb}{0.5,0.5,0.5}

\newcommand{\ig}{\textsf{IG}\xspace}
\newcommand{\tool}{\textsf{FORCE}\xspace}
\newcommand{\ifour}{\textsf{I4}\xspace}
\newcommand{\duoai}{\textsf{DuoAI}\xspace}
\newcommand{\flyvy}{\textsf{Flyvy}\xspace}
\newcommand{\clingo}{\textsf{Clingo}\xspace}
\newcommand{\folsep}{\textsf{FOL-IC3}\xspace}
\newcommand{\ictpo}{\textsf{IC3PO}\xspace}
\newcommand{\pdr}{\textsf{PDR/IC3}\xspace}
\newcommand{\distai}{\textsf{DistAI}\xspace}
\newcommand{\swiss}{\textsf{SWISS}\xspace}

\newcommand{\pfolsep}{\textsf{P-FOL-IC3}\xspace}
\newcommand{\scimitar}{\textsf{Scimitar}\xspace}
\newcommand{\popper}{\textsf{Popper}\xspace}

\newcommand{\ie}{\emph{i.e.}\xspace}

\newcommand{\eg}{\emph{e.g.}\xspace}

\newcommand{\etal}{\emph{et~al.}\xspace}

\newcommand{\cf}{\textit{cf.}\xspace}
\newcommand{\wrt}{\emph{w.r.t.}\xspace}

\definecolor[named]{ACMBlue}{cmyk}{1,0.1,0,0.1}
\definecolor[named]{ACMYellow}{cmyk}{0,0.16,1,0}
\definecolor[named]{ACMOrange}{cmyk}{0,0.42,1,0.01}
\definecolor[named]{ACMRed}{cmyk}{0,0.90,0.86,0}
\definecolor[named]{ACMLightBlue}{cmyk}{0.49,0.01,0,0}
\definecolor[named]{ACMGreen}{cmyk}{0.20,0,1,0.19}
\definecolor[named]{ACMPurple}{cmyk}{0.55,1,0,0.15}
\definecolor[named]{ACMDarkBlue}{cmyk}{1,0.58,0,0.21}
\definecolor[named]{PaleGreen}{RGB}{196, 255, 231}
\definecolor[named]{PaleOrange}{RGB}{255, 213, 169}
\definecolor{intnull}{RGB}{213,229,255}

\hypersetup{colorlinks,
  linkcolor=ACMDarkBlue,
  citecolor=ACMPurple,
  urlcolor=ACMDarkBlue,
  filecolor=ACMDarkBlue}

\setlist[itemize]{leftmargin=*}
\setlist[enumerate]{leftmargin=*}

\title{Inductive First-Order Formula Synthesis by ASP: \\A Case Study
  in Invariant Inference\thanks{We thank Andrew Cropper, Roland
    Kaminski, and Jianan Yao for providing help on \popper, \clingo, and \duoai, respectively. We are also grateful to Roland Yap for his
    feedback on a draft of this paper. This work was partially
    supported by a Singapore Ministry of Education (MoE) Tier 3 grant
    ``Automated Program Repair'' MOE-MOET32021-0001 and by Amazon
    Research Award ``Scaling Automated Verification of Distributed
    Protocols with Specification Transformation and Synthesis''.}}
\author{Ziyi Yang \qquad\qquad George Pîrlea \qquad\qquad Ilya Sergey
  \institute{National University of Singapore, Singapore} \email{yangziyi@u.nus.edu \qquad gpirlea@u.nus.edu \qquad
    ilya@nus.edu.sg} } \def\titlerunning{Inductive First-Order Formula
  Synthesis by ASP} \def\authorrunning{Z. Yang \etal}
\begin{document}
\maketitle
\begin{abstract}
  
We present a framework for synthesising formulas in first-order
logic (FOL) from examples, which unifies and advances
state-of-the-art approaches for inference of transition system
invariants.
To do so, we study and categorise the existing methodologies, encoding techniques in
their formula synthesis via answer set programming (ASP).
Based on the derived categorisation, we propose \emph{orthogonal
slices}, a new technique for formula enumeration that partitions the
search space into manageable chunks, enabling two approaches for
incremental candidate pruning.
Using a combination of existing techniques for first-order (FO)
invariant synthesis and the orthogonal slices implemented in our
framework \tool, we significantly accelerate a state-of-the-art
algorithm for distributed system invariant inference.
We also show that our approach facilitates \emph{composition} of
different invariant inference frameworks, allowing for novel
optimisations.


\end{abstract}

\section{Introduction}

%
First-Order Logic (FOL) has been used with great success as a
foundational tool for modelling and verifying complex systems. Its
applications span various domains, ranging from hardware design
\cite{een2011efficient} to software verification
\cite{Padon-al:PLDI16}.
These successes are largely attributed to the development of advanced frameworks that allow for automated verification and
synthesis, often supported by high-performance provers such as
Z3~\cite{Z3} and cvc5~\cite{BarbosaBBKLMMMN22}.

%
To achieve \emph{fully automated} verification of a complex system in
FOL, it is often necessary to \emph{synthesise} formulas capturing
the invariants (\ie, the properties always hold) of the system being verified.
Despite the undecidable nature of this task, many recent efforts have
made substantial progress to infer inductive invariants of complex
distributed systems (\eg, Lamport's
Paxos consensus protocol~\cite{Lamport98})
by synthesising FO formulas from examples: sampled traces of a
protocol or counter-examples to induction.
The resulting approaches are implemented by a plethora of distinct
frameworks~\cite{Yao-al:OSDI22,frenkel2024efficient}, and a systematic
study of their inter-connections is still missing.
%
This raises important questions: are the existing synthesis methods
fundamentally non-overlapping? Could techniques developed for one
approach be adapted to benefit others? Addressing these questions
would not only deepen our understanding of the underlying
methodologies but also enable the development of superior tools for
formula synthesis, potentially improving scalability of automated
verification tools across various domains.

%


In this paper, we present a unified framework for synthesising bounded
first-order formulas from examples---first-order structures that the
formulas satisfy. 
Our framework is designed to encode and \emph{combine} diverse
synthesis techniques, enabling seamless integration with different
high-level algorithms for invariant inference.
To achieve this, we have conducted a detailed study of \emph{nine}
recent approaches for  invariant inference of distributed systems (DS), each
of which offered a different take on inductive synthesis of
FO formulas.
We summarise our study in two main observations \wrt inductive
synthesis of FO formulas, hinting an opportunity for
improvement in the state of the art:
\begin{enumerate}[label=\textbf{O\arabic*}]
\item\label{o1} 
Existing techniques fall within a small number  of distinct classes, in terms of how
they treat their inputs and results, \eg, how examples are used and how
formulas are constructed. We give a uniform categorisation of these
approaches.

\item\label{o2} 
Whilst a wide variety of synthesis techniques exists, the vast
majority fall into one of two categories, exploiting the properties of
the first-order theories they employ: ``redundancy elimination'' and
``incremental pruning''.

\end{enumerate}

\noindent
To make a unified framework that captures~\ref{o1}, we use Answer Set
Programming (ASP) \cite{lifschitz2002answer}, to encode the enumeration-based FO formula synthesis (as constraint solving) and
 the customisations and techniques of the synthesis (as knowledges representation).
 To improve on the existing techniques, we exploit \ref{o2} by
 proposing the idea of \emph{orthogonal slices} (also implemented by
 ASP) of the FO search space: a new approach to efficiently prune
 candidate formulas during the inductive synthesis.
The key idea of orthogonal slices is to partition the search space into ordered
slices, where the synthesis of former slices can be used to prune the latter
slices using \emph{either satisfied or unsatisfied} formulas.
The practical benefits of the unified framework, \tool (\textbf{F}irst-\textbf{O}rder synthesiser via
  o\textbf{R}thogonal sli\textbf{CE}s), are demonstrated by improving two
state-of-the-art DS invariant synthesisers,
\duoai \cite{Yao-al:OSDI22} and \flyvy \cite{frenkel2024efficient},
\emph{without any conceptual modifications to their high-level
  algorithms}.
Our results show that our framework is sufficiently expressive and
extensible to encode and compose existing formula synthesis techniques,
advancing the state of the art in DS
invariant synthesis.

\section{Overview}
\label{sec:overview}

We start with a primer on inductive synthesis of FO
formulas---a common subroutine in existing invariant inference
frameworks for distributed systems~(DS). 
Then we summarise nine notable existing approaches
for DS invariant inference (with the earliest dated
2019), concluding with a brief description of our
ASP-based framework to capture various aspects of the synthesis
and its particular instance, orthogonal slices.

\subsection{Problem Definition}
\label{sec:overview:problem}

\paragraph{First-Order Language.}

In this work, we focus on system properties that are expressible in a
first-order, many-sorted logic with equality, following the common textbook definitions.
A \emph{signature} \(\Sigma = \langle C, R, F, S \rangle\) consists
of: 
a set of \emph{constant symbols} \(C\), a set of \emph{relation symbols} (predicates) \(R\), a set of \emph{function symbols} \(F\), and a set of \emph{sorts} \(S\) for the variables, constants, and function symbols.
In the rest of this paper we assume all signatures to be
\emph{finite}, \ie, the sets \(C\), \(R\), \(F\), and \(S\) are
finite.

Logic \emph{terms} are defined recursively. A term is either
%
a \emph{constant} \(c \in C\), a \emph{variable} \(x\), or a \emph{function symbol} \(f \in F\) applied to other terms (\eg, \(f(x_1, x_2)\)).
Logic \emph{atoms} are the basic formulas formed by applying
relation symbols from \(R\) or equality to terms of appropriate sorts;
and a \emph{literal} is an atom or its negation (\ie,
\(\neg p(x)\)). \emph{Formulas} are constructed by closing literals
under \emph{logical connectives} (\ie, conjunction \(\land\) and
disjunction \(\lor\)) and \emph{quantification} (universal
\(\forall\) and existential \(\exists\)). 
It is well-known that any FO formula can be transformed into
an equivalent formula in \emph{prenex normal form}, where all
quantifiers are placed at the beginning. In this structure, the
\emph{prefix} includes the quantifiers, and the \emph{matrix} 
consists of the remaining Boolean components. For example, the formula
\(\forall x : s_1 \, \exists y : s_2.~ p(x, y)\lor (\neg q(x) \land
r(y))\) is in prenex normal form, with prefix
\(\forall x : s_1 \, \exists y : s_2\) and matrix
\(p(x, y)\lor (\neg q(x) \land r(y))\).

    \begin{definition}[First-Order Synthesis Problem]
        \label{def:fosynthesis}
        Given a set of formulas $\Omega_0$ over signature \(\Sigma\) and a set of FO structures \(\sigma = \{ M_1, \dots, M_k \}\), where each \(M_i\) is a model over \(\Sigma\), find a set of formulas \(\Phi = \{\phi_1, \ldots, \phi_n\} \subseteq \Omega_0\) s.t.~\(\forall\phi \in \Phi\):
        \[
        \begin{aligned}
        &\text{1. } \forall M \in \sigma. M \models \phi, \quad &&\textsf{(satisfies all input FO structures)}, \\
        &\text{2. } \text{FreeVars}(\phi) = \emptyset, \quad &&\textsf{(closed formula)}, \\
        &\text{3. } \exists M \not\in \sigma. M \not\models \phi, \quad &&\textsf{(non tautology)}, \\
        &\text{4. } \forall \phi' \in \Phi. \phi \neq \phi' \land \phi \not\models \phi', \quad &&\textsf{(no formula entails another)}.
        \end{aligned}
        \]
    \end{definition}
In other words, the problem is to find in search space $\Omega_0$ a conjuncted set of well-formed
formulas~$\Phi$ that satisfy all the given first-order
structures in $\sigma$.
Such a conjunction describes the ``most precise'' formula that
satisfies all the given structures, because there is no satisfied formula which is entailed by any $\phi_i$. The problem can be seen as an instance of
the general specification synthesis problem~\cite{Park-al:OOPSLA23} in the setting of positive-only learning~\cite{sippy}.
An example of the problem is illustrated in \autoref{app:example}.

It is worth noting that we make several assumptions and
simplifications to the problem definition above:
(1)~a disjunction of all input FO structures is always a valid (though
overfit) solution, but in practice, meaningful formulas are to be found
in a size-restricted search space;
(2)~search spaces with function symbols and constants can be easily
handled by introducing new literals \cite[\S4]{Yao-al:OSDI21},
so we will avoid discussing them in detail; 
(3)~we assume users want to synthesise prenex DNF formulas, as DS invariants
are commonly expressed in this form; our techniques can be extended to
other FO formulas.

\subsection{Taxonomy of Existing Invariant Inference Algorithms}
\label{sec:existing}

Existing invariant inference methods use different formula enumeration
techniques and employ a variety of optimisations to effectively reduce the search
space.
We summarise nine representative existing approaches: 
\ifour~\cite{Ma-al:SOSP19}, \folsep~\cite{Koenig-al:PLDI20}, \ictpo~\cite{Goel-al:NFM21},
\swiss~\cite{Hance-al:NSDI21}, \distai~\cite{Yao-al:OSDI21},
\pfolsep~\cite{Koenig-al:TACAS22}, \duoai~\cite{Yao-al:OSDI22},
\scimitar~\cite{Schultz-al:CoRR}, and \flyvy~\cite{frenkel2024efficient} (in the order of their publication dates).

The categories are based on five detailed sub-aspects.
The first two sub-aspects are ``system-level'', capturing the
high-level design of a synthesis framework, while the remaining three
are ``algorithm-level'': they define the pruning techniques applied to
the brute-force enumeration of the search space.
%

\subsubsection{System-Level Aspects}
\paragraph{Inference mode.} 

This aspect determines the way the overall inference procedure uses FO
formula synthesis. In particular, \emph{one-shot} synthesis \cite{Ma-al:SOSP19,Yao-al:OSDI21,Yao-al:OSDI22} generates
all satisfied formulas in the search space given fixed input examples
(\eg, sampled traces of a distributed protocol), while
\emph{multi-shot} synthesis \cite{Schultz-al:CoRR,frenkel2024efficient} generates a set of candidate formulas
incrementally, based on examples obtained incrementally (\eg,
counter-examples of current invariants).
Combined approaches \cite{Koenig-al:PLDI20,Goel-al:NFM21,Hance-al:NSDI21,Koenig-al:TACAS22} use system traces or counter examples to guide the multi-shot
synthesis, calling the synthesis procedure multiple times.

Historically, the majority of multi-shot synthesis algorithms can be
seen as extensions of IC3~\cite{Jhala-Schmidt:VMCAI11}, while
one-shot synthesis can be considered as extensions of
Houdini~\cite{DBLP:conf/fm/FlanaganL01}. 
Crucially, both modes share the underlying FO synthesis problem similar to the definition of our synthesis problem,
which means our \tool framework can be used to improve
most existing approaches.


\paragraph{Language restriction.} 

To make the synthesis problem tractable, it is frequently defined for
a subclass of first-order logic.
The most common one is the Effectively Propositional Logic
(EPR~\cite{piskac2010deciding,Emmer-al:FMCAD10}) fragment, a subset of
FOL in which formulas can be transformed into equivalent propositional
formulas, allowing for provably decidable verification.

While EPR and its extensions are applied in most existing approaches with a theoretical decidability guarantee, in practice,
synthesis approaches often impose additional other syntactic
constraints. 
%
For example, the k-pseudo-DNF (proposed in
\pfolsep~\cite{Koenig-al:TACAS22}, and further applied to \cite{frenkel2024efficient}) is a syntax restriction for practical
efficiency: it is based on the observation that the invariant formulas
written in such form are smaller than standard DNF, reducing the search
space that needs to be explored. Specifically, a k-pseudo-DNF formula has the matrix in the form of $c_1 \rightarrow (c_2 \lor \ldots \lor c_k)$, where $c_i$ is a conjunction of literals. This is essentially a heuristic of the search, which makes sense because implications are commonly used to express invariants.

\distai~\cite{Yao-al:OSDI21} proposed
sub-templates (further applied to \cite{Yao-al:OSDI22}) for efficiency, exploiting the following property of
first-order logic with equality:
\begin{equation*}
    \forall X_1, X_2:T.\textit{mat}(X_1,X_2) \equiv  \forall X:T.\textit{mat}(X, X) \land \forall X_1 \neq X_2:T.\textit{mat}(X_1,X_2)
    \tag{0}
    \label{eq:mat-formula}
\end{equation*}
With such a property, formula synthesis can avoid enumerating formulas
on the LHS of \autoref{eq:mat-formula} because they are equivalent to the conjunction of
the two RHS formulas. Therefore, the only enumerations among two sub-templates on the right are required to synthesis the satisfied formula on the left, resulting in an accelerated enumeration.

More than syntactic constraints above, specific tools \cite{Koenig-al:PLDI20,Hance-al:NSDI21,Schultz-al:CoRR,frenkel2024efficient} also define their own syntactic customisations.
However, upon close examination of existing invariant inference
implementations, we find that syntactic constraints are not
extensible in many tools.
For example, extending k-pseudo-DNF with sub-templates would require
modifying any algorithm that manipulates formulas in \pfolsep.

\subsubsection{Algorithm-Level Aspects}

\paragraph{Redundancy elimination.} 
The most common pruning to apply is redundancy elimination, which is used in many synthesis tasks.
The idea is to simply eliminate the formulas that are equivalent to
each other. Approximations as ``pruning by symmetry'' (to be detailed in \autoref{sec:language})
are often used, where the symmetry of formulas
under quantification is identified to prune away redundant ones.
%
%
%
More than the syntactic-based equivalence, the redundancy can also be introduced by semantics (\eg,~tautology and contradiction), where many case-by-case rules are used in \distai~\cite{Yao-al:OSDI21} and \duoai~\cite{Yao-al:OSDI22}. An example of formulas eliminated by this kind of redundancy is the one containing $p(X) \land \neg p(X)$ or $p(X) \lor \neg p(X)$ as sub-expressions.
Another approach to redundancy is elimination \emph{canonicalisation},
which is implemented by \flyvy~\cite{frenkel2024efficient}: it also uses symmetry breaking, but defines a
partial order of formula sets (instead of the individual formulas) to eliminate redundancy.

A more complex but very effective pruning strategy of this category is
proposed in \duoai~\cite{Yao-al:OSDI22}, where a large number of DNF
formulas are shown to be redundant by their \emph{decomposition}: \(A \equiv B \land C\).
Similar to \autoref{eq:mat-formula}, the formula $A$ can be decomposed into
a conjunction of smaller formulas, so the original formula can be
pruned if all the smaller formulas are in the search space.
However, the authors of \flyvy~(\cf \cite[Appendix D]{frenkel2024efficient}) found that the decomposition in \duoai is
unsound in certain cases (\ie,~it leads to over-pruning), and proposed
an amended version.
We found it not easy to switch the \duoai
implementation to the amended version, as it uses the intermediate
results of the original decomposition in the overall synthesis loop,
which also makes it challenging to fairly compare the efficiency of the two
methods.

\paragraph{Incremental pruning.} 
Another inherent part of nearly all efficient approaches to FOL formula
synthesis is incremental pruning.
The idea is to test formulas against the set of input FO structures in a
specific order, exploiting the entailment relation between formulas to
eliminate the need to test some of them. 

The \emph{implication graph} to be detailed in \autoref{sec:implication-graph} is an effective technique of this aspect from \duoai~\cite{Yao-al:OSDI22}. More than this, \duoai~also identify another incremental pruning technique is based on
\emph{co-implication}, which combines the intermediate results of FO model checking with the redundancy elimination to prune the search space.
For example, if we find that formula $\forall x.P(x) \Rightarrow
R(x)$ satisfies all examples, we do not need to the test any formulas
of the form $\textit{prefix}.~(P(x) \land R(x) \land F) \lor G$,
since they are co-implied by the original formula and
$\textit{prefix}.~(P(x) \land F) \lor G$. 

\paragraph{Other techniques.}
Several pruning techniques are introduced in existing invariant inference tools~\cite{Koenig-al:PLDI20,Goel-al:NFM21,Schultz-al:CoRR}.
We will discuss them in \autoref{sec:related}, as they are not generally applicable to all FO synthesis tasks.


\vspace{1em}

To sum up the taxonomy of existing approaches, it suggests a possibility of a unified framework to capture
the existing techniques for FO synthesis and propose new ones.
In particular, we find that the \textbf{Language} and
\textbf{Redundancy} aspects together form a \emph{static} search
space, while the \textbf{Incrementality} aspect defines the
\emph{dynamic} pruning of the search space during the synthesis. 
These aspects are applicable to any FO synthesis problem.


\subsection{\tool: ASP-based Synthesis + Orthogonal Slices}
\label{sec:overview:tool}
\begin{figure}[t]
  \centering
  \includegraphics[width=\textwidth]{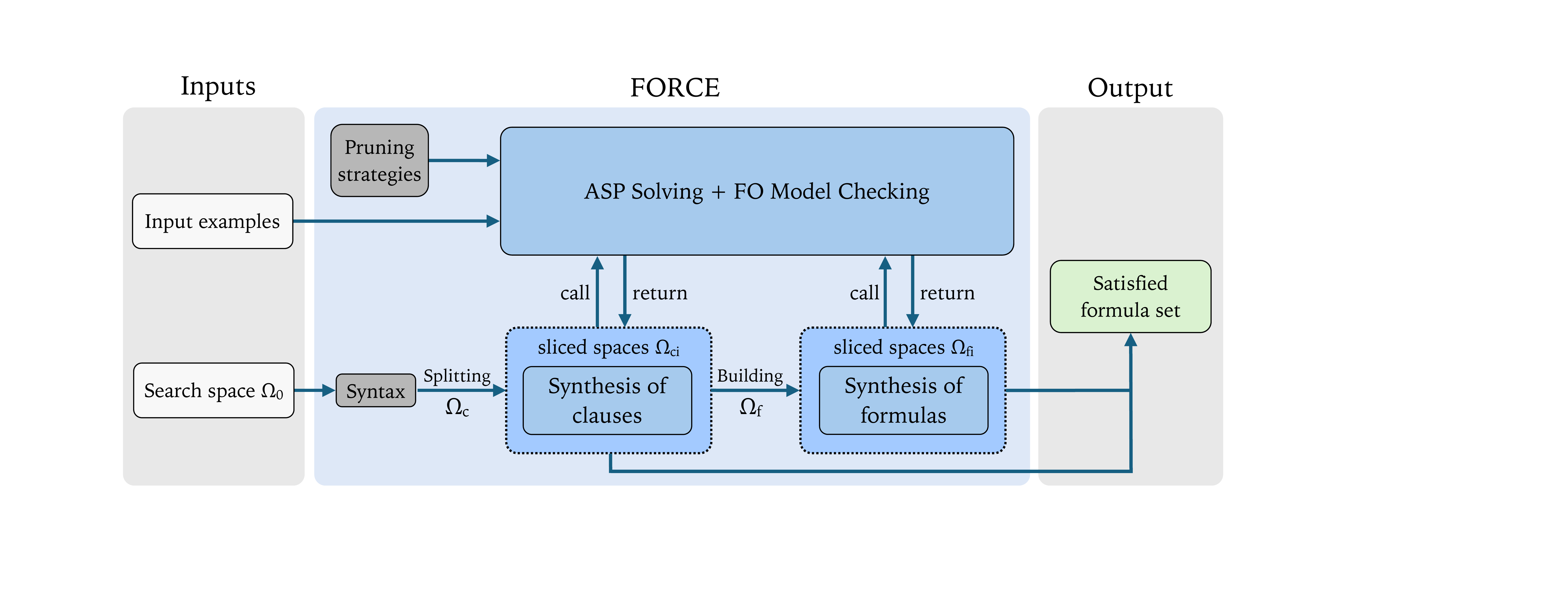}
  \setlength\abovecaptionskip{-5pt}
  \setlength\belowcaptionskip{-10pt}
  \caption{The workflow of \tool.}
  \label{fig:algo}
\end{figure}

Before the technical details, we give a high-level overview of
our ASP-based synthesiser \tool, whose workflow is shown in
\autoref{fig:algo}, by combining pruning strategies with the new \emph{orthogonal slices} technique.
The framework is customised by the grey parts: (1) the FO language (DNF
by default) and (2) a set of pruning rules (predefined but extensible).
Given the initial search space $\Omega_0$ and a set of FO
structures as input,
\tool starts by splitting the search space of clauses $\Omega_c$ from
$\Omega_0$. For further slices of $\Omega_c$ (called $\Omega_{c_1}$,
$\Omega_{c_2}$, ...), the algorithm generates the formulas in
$\Omega_{c_i}$ (by ASP solving), tests them on the examples (by
model checking), and prunes the later slices ($\Omega_{c_j},~j>i$)
based on the results.
Then using the output of $\Omega_c$, the algorithm further builds the
search space of formulas \emph{other than clauses} $\Omega_f$ into $\Omega_{f_1}$, $\Omega_{f_2}$, $\ldots$,
reusing the same process as in $\Omega_c$ to generate-test-prune.
Finally, \tool outputs both the clauses and the non-clause formulas that
satisfy the inputs. 
It is called \emph{orthogonal} because (1) slicing $\Omega_0$ into
$\Omega_c$ and $\Omega_f$ and (2) slicing of $\Omega_c$ and $\Omega_f$ are using different
pruning strategies and work together.


\section{Static Search Spaces of First-Order Formulas in ASP}
\label{sec:language}

We assume the reader is familiar with the basics of ASP, such as rules, choices, aggregates, and refer to the literature for more details~\cite{gebser2022answer}. This section demonstrates how ASP is suitable for encoding the static search space of FO formulas.

Our high-level approach follows the ``generate-and-test'' workflow, similar to other invariant synthesis systems, but differs in how the search space and pruning techniques are encoded. ASP is a paradigm suitable for such exhaustive enumeration with restrictions (search space customisations and pruning techniques in our case).
As shown in \autoref{fig:generic}, existing imperative approaches (demonstrated on the left) require manual integration of pruning steps (line 3-6), which can lead to brittle and hard-to-maintain code due to dependencies on intermediate results and evaluation order.
In contrast, our ASP-based approach on the right modularises the search space and pruning strategies directly into the solver (line 2). This ASP-based ``generate-and-test'' loop improves extensibility and maintainability, by allowing easier integration of new pruning rules (as new domain knowledge) and automatic handling of dependencies.
%
In the remainder of this section, 
we first demonstrate how to encode a basic search for FO formulas and then show the search is customised with different knowledge of pruning.

\begin{figure}[t]
    \centering
    \small
    \begin{minipage}[t]{0.45\textwidth}
        \begin{algorithmic}[1]
            \State $\textit{tmp} \gets \textit{init}()$
            \ForAll{$x \in \Omega_0$}
                \If{$\textit{pruning1}(x, \textit{tmp})$}
                    \State \textbf{continue}
                \ElsIf{$\textit{pruning2}(x, \textit{tmp})$}
                    \State $\textit{update2}(\textit{tmp}, x)$
                \Else
                    \If{$\textit{Satisfied}(x)$}
                        \State \textit{tmp.update}($x$)
                    \EndIf
                \EndIf
            \EndFor
            \Return $\textit{tmp.result}$
        \end{algorithmic}
    \end{minipage}
    \hspace{0.05\textwidth} 
    \begin{minipage}[t]{0.48\textwidth}
        \vspace{-2em}
        \begin{algorithm}[H]
            \caption{Extending \autoref{def:fosynthesis} with Customised Pruning Rules}
            \label{alg:extended-fosynthesis}
            \begin{algorithmic}[1]
                \Function{FOSyn}{$\Omega_0$, $\sigma$, \textit{prunings}}
                \State $\textit{solver} \gets \textit{init}(\Omega_0, \textit{prunings})$
                \While{$\textit{solver.solve}()$}
                    \State $x \gets \textit{solver.get\_current}()$
                    \If{$\textit{Satisfied}(x)$}
                        \State \textit{yield} $x$
                    \EndIf
                \EndWhile
                \EndFunction
            \end{algorithmic}
        \end{algorithm}
        
    \end{minipage}

    \caption{The synthesis loop (imperative programming vs. ASP solving).}
    \label{fig:generic}
\end{figure}


\subsection{Encoding the Enumeration}
\label{sec:encoding-enum}

Let us show how to encode FO formula enumeration in ASP.
As an illustration, we use \duoai's~\cite{Yao-al:OSDI22} configuration of FO search space for
synthesising invariants of the \texttt{lockserv} distributed protocol. 

{\footnotesize
\begin{verbatim}
  var: node: n1, n2; lock: l1
  relations: lock_msg:          node, lock;  grant_msg:  node, lock; 
             unlock_msg:        node, lock;  holds_lock: node, lock; 
             server_holds_lock: lock
  max-literal: 4 max-or: 3 max-and: 3 max-exists: 1
\end{verbatim}
}

\noindent
The first two lines, ``var'' and ``relations'', together specify the
variables (in \texttt{node} and \texttt{lock} sorts) and atoms (made out of
five relations and valid variables) that can be used in the formulas.
Note that the order of variables in formulas' prefix is usually fixed in EPR formulas,
\ie, the variable \verb|l1| cannot appear before any node
variable (\texttt{n1}, \texttt{n2}) in the prefix. 
The search space of formulas in DNF is then constrained by the
problem-specific customisations: the maximum number of literals in the
formula, the maximum number of disjunctions, the maximum number of
cubes (\ie, conjunctions of literals), and the maximum number of
existential quantifiers.
They together define the $\Omega_0$ in \autoref{def:fosynthesis}.

Thanks to the simplicity of FO formulas' prenex normal form, the enumeration
 can be easily encoded in ASP. The enumeration of
formulas requires generating their two parts: prefixes and matrices, within the 
restrictions of the search space. The ASP encoding of the basic search space is illustrated as follows:
%

{\footnotesize
\begin{verbatim}
    var(node, n1). var(node, n2). var(lock, l1).             % variables
    0{exists(Var): var(Var, _)}1.                            % prefix bounded by max-exists
    rel(lock_msg, (node, lock)). ...                         % relations
    vars((node, lock), (n1, l1)). ...                        % variable tuples
    atom(Pred, Args) :- rel(Pred, Types), vars(Types, Args). % atoms
    pos(0..1). cube(1..3).                                   % sign of literal, max-or
    0{lit_in_C(P,A,Pos,C):atom(P, A), pos(Pos)}3 :- cube(C). % matrix bounded by max-and
    :- #count{P,A,Pos,C: lit_in_C(P,A,Pos,C)} >= 4.          % max-literal
\end{verbatim}
}

\noindent
In the program above, the prefix and matrix generation of formulas are achieved by the choice construct \verb|0{...}n| on \verb|exists()| and \verb|lit_in_C()| predicates, which are restricted by the parameters in the configuration. 
The last line then eliminates the answer
sets where more than four literals are in the corresponding formula.
As an example, the answer set $\{$\verb|lit_in_C(lock_msg,(n1,l1),0,1)|,
\verb|lit_in_C(grant_msg,(n2,l1),1,1)|,
\verb|lit_in_C(unlock_msg,(n1,l1),0,2)|, \verb|exists(l1)|$\}$
corresponds to $\forall \texttt{n1~n2}, \exists \texttt{l1}.~(\neg\texttt{lock\_msg(n1,l1)}\land\texttt{grant\_msg(n2,l1)}~)\lor\neg\texttt{unlock\_msg(n1,l1)}$. We should also note that the encoding provided here is to illustrate the basic idea of the enumeration; for different syntactic customisations (\eg, the variable order in the prefix) of FOL formulas, certain predicates and rules are required to be added to the encoding.


\subsection{Encoding Pruning by Redundancy}
\label{sec:encoding-redundancy}

A particular encoding of a search space can output many answer sets
whose corresponding formulas not necessarily satisfy the examples or
even basic well-formedness constraints.
For instance, the formula
$\forall n_1. \, \texttt{lock\_msg}(n_2, l_1)$ should be ignored
because $n_2$ is not in the prefix.
As another example,
$\textit{prefix}. \, \texttt{lock\_msg}(n_1,
l_1)\lor\texttt{lock\_msg}(n_2, l_1)$ is equivalent to
$\textit{prefix}. \, \texttt{lock\_msg}(n_2,
l_1)\lor\texttt{lock\_msg}(n_1, l_1)$ but can be featured twice as two
different answer sets. 
Our next step is, therefore, to encode those \textbf{Redundancy}
techniques from \autoref{sec:existing} on the search space to only
output well-formed formulas.

Let us illustrate the \emph{symmetry-based normalisation} technique
allowing to exploit equivalence using ASP; the remaining encoding of
redundancy elimination can be found in our implementation. The
normalisations are done by building a partial order on the formulas.
That is, we can make sure that (1) the number of literals in cubes
(from left to right in a DNF) is non-decreasing,
(2) the minimal (in alphabetic order) literal in a cube $i$ is less
than the minimal literal in cube $j>i$ if their number of literals are
the same,
and (3) if two variables of one sort $v_i<v_j$, then the minimum predicate
where $v_i$ appears should be less or equal to the minimum predicate where $v_j$
appears.
The following ASP program encodes these pruning rules by constraining
the orders in the formula.

{\footnotesize
\begin{verbatim}
  lit_no_in(No, C) :- lit_in_C(P,A,Pos,C), lit_no(P,A,Pos,No).
  num_lit(C, N) :- cube(C), #count{P,A,Pos: lit_in_C(P,A,Pos,C)} = N.
  :- num_lit(C1,N1), num_lit(C2,N2), C1 < C2, N1 > N2.
  min_lit(C, Min) :- cube(C), #min{No:lit_no_in(No,C)} = Min.
  :- num_lit(C1,N), num_lit(C2,N), min_lit(C1,Min1), min_lit(C2,Min2), C1<C2, Min1>Min2.
  min_lit_var(V, Min) :- used_var(V), #min{P:lit_in_C(P,A,_,_), var_in(V,A)} = Min.
  :- min_lit_var(V1,Min1), min_lit_var(V2,Min2), V1 < V2, Min1 > Min2.
\end{verbatim}
}

\noindent
To summarise this section, we have shown that for the synthesis of
FO formulas, the existing techniques can be easily
encoded in ASP, which works as the backbone of our synthesiser.
%
The expressive power of ASP allows us to encode existing techniques
shown in \autoref{sec:existing} concisely and make them
work together efficiently, being further optimised with our new
pruning technique, described in the next section.
%



\section{Dynamic Search Spaces via Orthogonal Slices}
\label{sec:sss}

Armed with the ASP-based framework formula enumeration in the previous section, which
allows one to combine pruning techniques, we propose \emph{orthogonal slices} to handle the dynamic search space, which (1) generalises
the state-of-the-art incremental pruning technique \emph{implication graph} (\ig), and (2) introduces a novel complementary pruning to resolve
the bottleneck of \ig. Both are achieved by slicing the FO search space into smaller ordered parts, and easily implemented by ASP's incremental solving.

\subsection{Implication Graphs}
\label{sec:implication-graph}
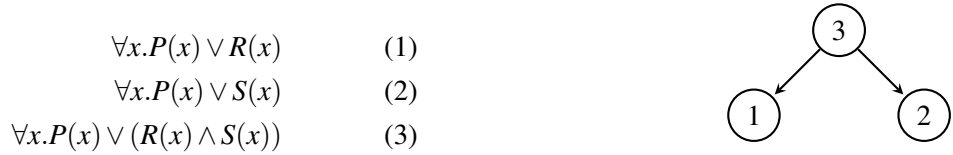
\begin{figure}[t]
    \centering
    \begin{minipage}{0.45\textwidth}
        \centering
        \begin{align}
    \forall x. P(x) \lor R(x) \label{eq:formula1} \\
    \forall x. P(x) \lor S(x) \label{eq:formula2} \\
    \forall x. P(x) \lor (R(x) \land S(x)) \label{eq:formula3} 
        \end{align}
    \end{minipage}
    \hfill
    \begin{minipage}{0.4\textwidth}
        \centering
        \begin{tikzpicture}[->,>=stealth,shorten >=1pt,auto,node distance=1.6cm,thick]
            \tikzstyle{every state}=[fill=white,draw=black,text=black,minimum size=10pt]

            \node[state] (F3) {3};
            \node[state] (F1) [below left of=F3] {1};
            \node[state] (F2) [below right of=F3] {2};

            \path (F3) edge node {} (F1);
            \path (F3) edge node {} (F2);
        \end{tikzpicture}
    \end{minipage}
    \caption{Left: Formulas example. Right: Implication graph of the formulas.}
    \label{fig:formulas-and-implication-graph}
\end{figure}

Amongst the proposed pruning techniques for distributed system
invariant synthesis in~\autoref{sec:existing}, \emph{implication
  graph} of FO formulas is a prominent one. 
An \ig is a directed graph where each node represents a
formula, and an edge from $A$ to $B$ indicates that $A$ implies $B$.
The pruning is processed by removing formulas that are implied by
already satisfied formulas from the search space, dynamically, to
accelerate the synthesis process. That said, we can call it ``pruning
by satisfied formulas''.
The \ig-based tool \duoai~\cite{Yao-al:OSDI22} can infer inductive invariants
for many complex protocols where no other existing tools can.

\begin{example}[An Illustration of the Implication Graph Shown in \autoref{fig:formulas-and-implication-graph}]
    Formula~\eqref{eq:formula3} implies both formula~\eqref{eq:formula1} and formula~\eqref{eq:formula2}. If formula~\eqref{eq:formula3} is satisfied, then formulas~\eqref{eq:formula1} and~\eqref{eq:formula2} are automatically satisfied and can be pruned from the search space.
\end{example}


\subsection{From the Formula to the Search Space}
\label{sec:slicing-tem}
The first step of the orthogonal slices is to \emph{abstract} the entailment relation from formulas to the level of the search spaces,
then synthesise formulas following the partial order on the search spaces to achieve incremental pruning. 
We represent a FO formula search space by means of imposing syntactic constraints of the
candidates, with the sets of possible values as arguments. As such, the search space is the Cartesian product of those sets, where
each element describes a sub search space.

\begin{definition}[Template of First-Order Formulas and Its Slicing]
    A template \( T \) of first-order formulas is defined by a tuple of parameter sets \( T = (P_1, \ldots, P_n) \), where each \( P_i \) represents a set of possible values for a parameter. The search space \( \Omega(T) \) defined by \( T \) is the Cartesian product of these parameter sets.
    A valid slicing of \( T \) (called \(\mathtt{SL}(T)\)) is defined as a partitioning of \( T \) into sliced-templates \( \{T_1, \dots, T_j\} \), where each sliced-template \( T_i = (P_{i1}, \ldots, P_{in}) \) corresponds to a subset of the parameter sets. 
    The search space \( \Omega(T) \) is then partitioned into  subsets \( \{\Omega(T_1), \dots, \Omega(T_j)\} \), where:
    \[
        \Omega(T) = \prod_{i=1}^n P_i. \qquad \bigcup_{i=1}^j \Omega(T_i) = \Omega(T), \quad \text{and} \quad \Omega_m \cap \Omega_n = \emptyset \quad \text{for all } m \neq n.
    \]
    This ensures that the entire search space is covered, without
    overlap between slices, and each slice \( \Omega_i \) corresponds
    to a sliced-template \( T_i \).
\end{definition}

\begin{example}[A Template of \texttt{lockserv} in \autoref{sec:encoding-enum}]
    \label{ex:template}
    
    \begin{itemize}
        \item The number of existential quantifiers $ne$,
        \item The number of variables used in each sort in a tuple  $tv~=~(n_{v1},~\ldots,~n_{vi},~\ldots)$,
        \item The number of literals in each cube sorted in a tuple  $tl~=~(n_{l1},~\ldots,~n_{li},~\ldots)$.
    \end{itemize}
    \noindent
    And the parameters of the template (after the redundancy pruning  in \autoref{sec:encoding-redundancy}) are:
    \begin{itemize}
        \item $P_{ne}=\{0,1\}$
        \item $P_{tv}=\{(0,1), (1,0), (1,1), (2,0), (2,1)\}$
        \item $P_{tl}=\{(1),(2),(3),(1,1),(1,2),(1,3),(2,2),(1,1,2)\}$
    \end{itemize}
\end{example}

Now we describe the slicing for ``pruning by satisfied formulas'', which is defined by the partial order of
the sliced-templates (and their parameters).

\begin{definition}[Partial Order for Parameters of Templates]
Given two parameters \( P_{ix} \) and \( P_{iy} \) as subsets of \( P_i \) in a template \( T = (P_1, \ldots, P_i, \ldots, P_n) \), we say that \( P_{ix} \) is less or equal than \( P_{iy} \) (denoted as \( P_{ix} \preceq P_{iy} \); \( P_{ix} \prec P_{iy} \) in case \(P_{ix} \neq P_{iy}\)) \wrt $\mathtt{SL}(T)$ if
\begin{align*}
    \forall T_1 = (P_1', \ldots, P_{ix}, \ldots, P_n'), &~T_2 = (P_1', \ldots, P_{iy}, \ldots, P_n') \in \mathtt{SL}(T), \notag \\
    \forall \phi \in \Omega(T_1), &~\exists \phi' \in \Omega(T_2) \text{ such that } \phi' \models \phi.
\end{align*}
\end{definition}

\begin{definition}[Partial Order for Sliced-templates]
    Given two sliced-templates \( T_i = (P_{i1}, \ldots, P_{in}) \) and \( T_j = (P_{j1}, \ldots, P_{jn}) \) sliced from $\mathtt{SL}(T)$, we say that \( T_i \) is less or equal than \( T_j \) (denoted as \( T_i \preceq T_j \)) if 
    $\forall k \in [1,n], P_{ik} \preceq P_{jk}$,
    where the equality holds if and only if \( P_{ik} = P_{jk} \) for all \( k \in [1,n] \).
\end{definition}

With two definitions above, the whole search space of FO formulas is sliced (and ordered) by the partial order; we call this procedure \textsc{SplitTem}. Note that possibly the search space's parameters are not as regular as in \autoref{ex:template}, but the worst case of the partial order is exactly the implication graph: the set of possible formula candidates is the only parameter, and the partial order is defined by FO entailment. This says, the parameterisation of FO search space by syntactic constraints is ``always'' possible.

\begin{example}[Partial Order of the Templates in \autoref{ex:template}]
    The partial order of the templates' parameters of \texttt{lockserv} is defined as follows:
    \begin{itemize}
        \item For $n_e,$ it follows the integer order: $\{0\} \prec \{1\}$.
        \item For $t_v,~\text{the set}~\{(n_{v1}, n_{v2}, \ldots, n_{vi}, \ldots)\} \prec \{(n_{v1}, n_{v2}, \ldots, n_{vi}+1, \ldots)\}$
        \item For $t_l,~\text{the set}~\{(n_{l1}, n_{l2}, \ldots, n_{li}, \ldots)\} \prec \{(n_{l1}, n_{l2}, \ldots, n_{li}-1, \ldots)\}$ and\\ $\{(n_{l1}, n_{l2}, \ldots, n_{li}, \ldots)\} \prec \{(n_{l1}+1, n_{l2}, \ldots, n_{li}, 1)\}$
    \end{itemize}
\end{example}

The partial order on template parameters is defined consistently with
formula entailment relations: by swapping $\forall$ into $\exists$, a
formula becomes more general; by replacing one variable with a fresh
one (together with the formula decomposition, proven in \cite[\S
4]{Yao-al:OSDI21}), a formula becomes more general; by deleting a
literal from a cube or adding a new cube, a formula becomes more
general. Therefore, by obtaining the partial order of a slicing of the
search space, the ``pruning by satisfied formulas'' is naturally
achieved.


%
The ASP encoding of ``pruning by satisfied formulas'' is done by multi-shot solving \cite{gebser2019multi}, which is standard to achieve incremental solving in ASP. 
The sketch of its encoding is as follows:

{\footnotesize
\begin{verbatim}
    #program inv(prefix,matrix).
    :- output(Prefix, Matrix), pre_weaken(Prefix, prefix), mat_weaken(Matrix, matrix).
\end{verbatim}
}
\noindent
which says that the formula entailed by the \texttt{inv} (\ie, a satisfied formula) should not be generated. The entailment checking is implemented by variable substitution together with prefix and matrix weakening (detailed in our implementation).

\subsection{Slicing DNF Modulo Clauses}
\label{sec:slicing-dnf}

The problem of implication graph (or ``pruning by satisfied formulas'' in general) is its scalability: the search starts from the root to leaves of the graph (top to bottom in
\autoref{fig:formulas-and-implication-graph}), but the number of root
nodes is still exponential in the search space's size. To see the issue, 
let us take the formula
$\textit{prefix}.~(\textit{lit}_{11} \land \textit{lit}_{12} \land
\textit{lit}_{13}) \lor (\textit{lit}_{21} \land \textit{lit}_{22})$ as an example:
it can be a root node used to prune the formula
$\textit{prefix}.~\textit{lit}_{1i} \lor \textit{lit}_{2j}$ because of the entailment.
However, checking all root formulas has complexity of $O(n^5)$ ($n$ is
the number of literals), but this effort prunes only formulas using
$O(n^2)$ space.

Intuitively, if ``pruning by satisfied formulas'' is costly when using
the result of a larger search space to prune the smaller one, its dual
version--``pruning by unsatisfied formulas'', should solve it.
This idea, however, is not immediately applicable for two reasons:
(1)~the two pruning strategies have different directions
(general-to-specific v. specific-to-general), which means an algorithm
needs to deal with both, and (2)~for complex problems, the majority of
formulas are unsatisfied (they do not cover all examples), which means
reducing the search space too many times incurs a large performance
overhead.

Our solution to
the first issue is simply another slicing, which is 
orthogonal to the slicing of templates:
first synthesising clauses (disjunctions of
literals) by slicing them from the whole search space (named \textsc{SplitDNF}), and then pruning the search of DNF synthesis based on the unsatisfied clauses.
To further address the second issue, we avoid the high-cost of ``pruning by
unsatisfied clauses'' by \emph{constructing} the DNF search space from the
satisfied clauses.
The definition of DNF search space construction given below
describes the pruning of this slicing.

%

\begin{definition}[Possibly satisfied DNF modulo clauses]\label{def:dnf-modulo-clauses}
    Given a set of satisfied (\wrt input examples) clauses $\Phi_c$, a
    formula in DNF of the form
    $ \textit{prefix}.~(\textit{lit}_{11}\land \ldots \land
    \textit{lit}_{1k_1}) \lor \ldots \lor (\textit{lit}_{m1}\land \ldots
    \land \textit{lit}_{mk_m}) $ is possibly satisfied only if
      $$
      \forall l_i \in \bigcup_{i=1}^{m} \{\textit{lit}_{i1}, \ldots, \textit{lit}_{ik_i}\},~\textit{prefix}.~l_1 \land \ldots \land l_m
      $$
      is satisfied \wrt $\Phi_c$. The function constructing DNFs is denoted as \textsc{BuildfromClauses}.
  \end{definition}

%
In plain words, a formula in DNF is added into the search space only when
all clauses it entails satisfy all input FO structures. The reason we resolve the bottleneck of ``pruning by satisfied formulas'' is evident: the search space of clauses, which is exponential
in the number of cubes, is much smaller than the search space of DNF
(exponential in the number of literals) for most cases. 
Moreover, since this search space construction is also naturally encoded by ASP, all existing techniques to formulas in \autoref{sec:existing} are
directly applied to further refine this search space built from
clauses.


\begin{table}[t]
    \centering
    \begin{tabular}{|l|l|l|}
        \hline
        \textbf{Synthesis} & \textbf{ASP} & \textbf{Connection} \\ \hline
        Parameterised search space & choice constructs & set Cartesian product \\ \hline
        Slicing & external statement & set partition \\ \hline
        Static pruning & integrity constraint & set comprehension \\ \hline
        Dynamic pruning & multi-shot solving & set union, difference \\ \hline
    \end{tabular}
    \caption{The connection between synthesis and ASP.}
    \label{tab:connection}
\end{table}

\pagebreak
\subsection{The Synthesis Algorithm}

\begin{wrapfigure}{r}{0.54\textwidth} 
    \vspace{-24pt} 
    \begin{minipage}{0.54\textwidth}
        \begin{algorithm}[H]
            \small
            \caption{The Core Algorithm of \tool}
            \label{alg:synthesis}
            \begin{algorithmic}[1]
                \Function{\tool}{$\Omega_0$,~$\sigma$,~$\textit{prunings}= \textit{pre\_def}$}
                    \State $\Phi_c,~\Phi_f \gets \emptyset$
                    \State $\Omega_c\gets \Call{SplitDNF}{\Omega_0}$ 
                    
                    \For{$\Omega_{ci} \in \Call{SplitTem}{\Omega_c}$}
                        \State $\Phi_i \gets \Call{FOSyn}{\Omega_{ci},~\sigma,~\textit{prunings}}$
                        \State $\Phi_c \gets \Phi_c \cup \Phi_i$
                        \State $\textit{prunings} \gets \textit{prunings.update}(\Phi_i)$
                    \EndFor
                    \State $\Omega_f \gets \Call{BuildfromClauses}{\Phi_c,~\Omega_0}$ 
                    \For{$\Omega_{fi} \in \Call{SplitTem}{\Omega_f}$}
                        \State $\Phi_i \gets \Call{FOSyn}{\Omega_{fi},~\sigma,~\textit{prunings}}$ 
                        \State $\Phi_f \gets \Phi_f \cup \Phi_i$ 
                        \State $\textit{prunings} \gets \textit{prunings.update}(\Phi_i)$
                    \EndFor
                    \State $\Phi_c \gets \Call{FilterImplied}{\Phi_c,~\Phi_f}$ 
                    \State \Return $\Phi_c \cup \Phi_f$
                \EndFunction
            \end{algorithmic}
        \end{algorithm}
    \end{minipage}
    \vspace{-12pt} 
\end{wrapfigure}
Given the formula synthesis algorithm in
\autoref{alg:extended-fosynthesis} for an input search space,
and the two approaches to slice the search space,
the procedure for synthesising formulas is
given by Algorithm~\ref{alg:synthesis}: first synthesise clauses by
its sliced template (lines~4 to~7), then synthesise other formulas
by the sliced template of DNF modulo clauses (lines 9 to 12), finally
normalising the satisfied formulas and output (lines 13 and 14). 
The input \textit{pruning} rules are customisable, but we pre-defined
existing redundant and incremental rules in
\autoref{sec:existing}.
The soundness of the algorithm \wrt orthogonal slices (\ie, it does
not over-prune) is guaranteed by the fact that if a formula $A$ is
pruned, $A$ is either unsatisfied, having being pruned by unsatisfied
clauses in DNF modulo clauses, or satisfied but more general than a
satisfied formula which prunes $A$ in a sliced template.

Looking back to the whole synthesis process, both static and dynamic search spaces we discussed are essentially achieved by different set operations, which is the algorithmic reason for us to use ASP. In \autoref{tab:connection}, we summarise the connection between the synthesis problem and ASP by unifying them as set operations, which hopefully helps the readers from either synthesis or ASP background to understand ``why and what synthesis task is suitable for ASP''.



\vspace{-10pt}

\section{Experimental Evaluation}
\label{sec:eval}

\subsection{Implementation and Settings}
Our implementation of \tool combines the use of ASP (250 lines for the
enumeration plus 80 lines for pruning, which is extremely concise to
encode the different existing and new pruning techniques), and C++ to call APIs
of \clingo~\cite{gebser2022answer} (400 lines for the synthesiser call
functions, and 200 lines to integrate with other systems).
The
benchmarks are run on a MacBook Pro with 8-core M1~Pro CPU,
16GB~RAM.  
The current version of \tool
is available at \url{https://zenodo.org/records/15654427}.

\label{sec:eval:impl}
\begin{figure}[t]
    \centering
    \includegraphics[width=\textwidth]{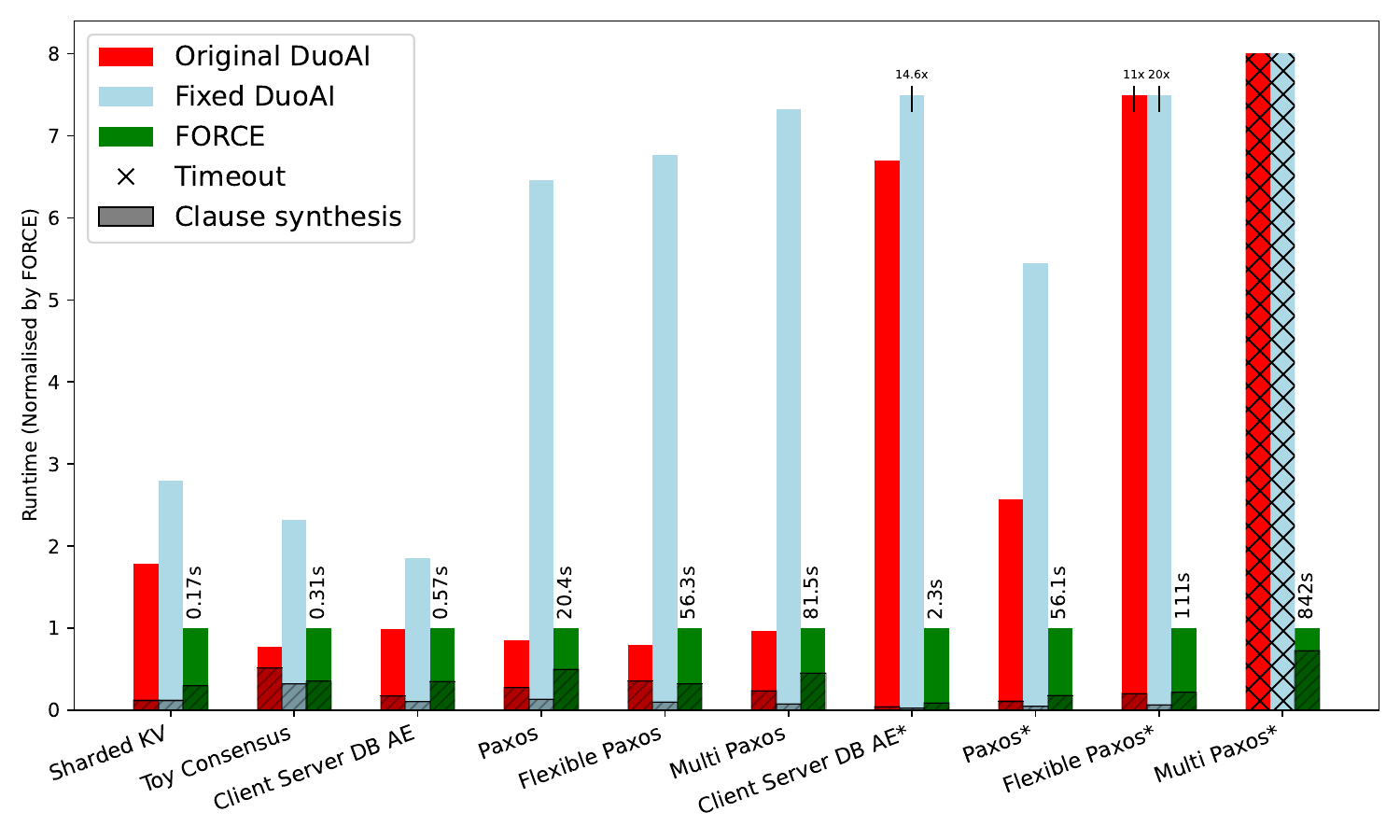}
    \caption{Invariant synthesis. Results are normalised \wrt \tool
      run times.}
    \label{fig:benchmark}
\end{figure}

We ran \tool on a set of benchmarks from the work
on \duoai~\cite{Yao-al:OSDI22}, the most efficient state-of-the-art
DS invariant inference tool.
Briefly speaking, \duoai's algorithm
is an enumeration-based inductive synthesis using sampled traces of the protocols as examples, followed by an optimised
Houdini~\cite{DBLP:conf/fm/FlanaganL01}. 
Since the formula enumeration component has non-negligible run time
(detailed later), reducing the overhead of
the enumeration part can significantly improve the overall
performance.

Note that (at least) two unsound optimisations were made in \duoai:
the first is an \emph{over-pruning}, found and rectified by~\cite{frenkel2024efficient}: \duoai identifies formulas that
can be decomposed into smaller formulas, then not testing the original
formula to reduce the search space; however, a subset of those should not be
decomposable, which leads to unsoundness by missing possibly satisfied
formulas.
The second unsound optimisation is the use of \emph{restricted quantifiers}  of formulas;
specifically, the default setting of \duoai only allows arbitrary
quantifiers over the last three variables in a template (\eg, for
$Q_1X_1.Q_2X_2.Q_3X_3.Q_4X_4$, $Q_1$ can only be~$\forall$), which
results in an incomplete search in general, but practically works for
their benchmarks.
We treat the first issue as a bug, and the second observation as
domain-specific knowledge that encodes an extra constraint on the
search space.
\vspace{-10pt}
\subsection{Results and Analysis}
Our performance statistics are shown in \autoref{fig:benchmark}. We
selected six complex distributed protocols from the \duoai suite
for evaluation on synthesising inductive invariants.
\tool was given the same input as \duoai: traces as input examples and
a search space configuration which contains the inductive invariant.
For each protocol, we benchmark with two configurations: one with the
restricted quantifier limitation from \duoai and one without it. 
The unrestricted quantifier setting is indicated by a $*$ in the figure
when enabled. The reason for showing only 10 data points, rather
than~12, is that the difference of quantifier restriction did not
affect the small search space for the first two smaller protocols. We
fixed the over-pruning bug in \duoai (referring to the result as
``fixed \duoai'') with our best effort for a fair comparison. That
said, we also provide results of the ``original'' \duoai version as a
reference point.

The results show that \tool significantly outperforms fixed \duoai in
all benchmarks, and also beats the original (unsound) \duoai in most
cases. The difference in performance is larger without the
quantifier restriction. 
As a reference, the original \duoai's runtime to synthesise invariants
for Paxos, FlexiblePaxos and MultiPaxos are 60.4s, 78.7s and 1,549s,
respectively; this means the enumeration in \duoai is the main
bottleneck for complex protocols, so the improvement is effective for overall runtime.

The improvement is mainly sourced from the effective pruning by DNF
modulo clauses (\autoref{sec:slicing-dnf}): as the \emph{clause
  synthesis} part in the bars shows, \tool takes (slightly) longer
time on the clause synthesis than \duoai, and results in a much
shorter time for the remaining synthesis by reducing the search space.
The huge difference for cases without quantifier restriction is also
explained for the same reason: taking Paxos as an example, the number of
satisfied clauses increased from 134 to 141 when disabling the restriction
in \tool, which means the increased search space for DNF is not much (with 36s
difference of time); but for \duoai, the enumeration explores the
whole new search space (thus 76s difference in total). Note that
abstracting the formula into the search space as in \autoref{sec:slicing-tem} allows
parallelism of the formula synthesis in \tool, which also
contributes to the efficiency, but it is specific to (1) the
implementation of parallelism, and (2) the multi-threaded solving in
\clingo, which is not essential to discuss here.
The overall extra time when disabling the parallelism in our system
varies from 50\% to 100\%, which shows potential improvement with a
better implementation of parallelism.

Another interesting comparison \wrt to the unsoundness in \duoai is
shown between the red and blue bars: the fixed version takes a longer
time to explore the (now larger) search space, but the clause
synthesis time is lower, because additional discovered satisfied
formulas prune the clauses entailed (which also implies the unsoundness
of original \duoai).
By analogy with our orthogonal slices, they are using results from
``larger slices'' to prune the ``smaller slices'', which results in 
inefficiency, as expected (see \autoref{sec:slicing-dnf}).
From a high-level aspect, this comparison illustrates the common
trade-off between completeness and efficiency in synthesis tasks, but
\tool achieves complete search with even better efficiency comparing to
the original \duoai by spending minor extra time (in the green bars) to synthesise
clauses.
%
%
\vspace{-10pt}
\subsection{On Composability of \tool}
\label{sec:eval:combination}


More than the performance benefits of the orthogonal slices, the ``solver-aided''  feature of \tool is also promising:
it works as a general framework to combine different approaches by providing
a common interface for FO synthesis. To illustrate this, we built the bridges for both \duoai
and \flyvy~\cite{frenkel2024efficient} (about 200 lines of code for each), and used the clause
output (\ie, part of \autoref{alg:synthesis} before line~8) to
optimise \flyvy.

The high-level task of \flyvy is to output the \emph{strongest}
inductive invariant (that contains a set of FOL formulas) given a
distributed system protocol and a bounded FO language.
Without describing the details of \flyvy's algorithm,
it is clear that if \flyvy is obtaining formulas, those formulas
can be bounded by DNF modulo clauses. We use the traces
obtained from \duoai together with the bounded language to synthesise
satisfied clauses and output them as an ASP program by \tool, then restricting
the formulas in \flyvy by calling the ASP program to filter
out the unnecessary formulas. The results show that, for two most complex
protocols can be synthesised in \flyvy, Paxos and FlexiblePaxos, the
sizes of the output invariants are reduced by about 13\% (165/1260) and 7\% (61/816) respectively
on average (detailed in \autoref{app:statistics}), and they remain strongest in the sense that
unnecessary formulas are removed during the intermediate steps of
\flyvy.
We believe it is a promising direction to explore different combinations of
existing approaches with the power of \tool.

\vspace{-10pt}

\section{Related Work and Discussion}
\label{sec:related}

\vspace{-5pt}
\paragraph{Inductive Logic Programming.}

Our work is closely related to the field of Inductive Logic
Programming~(ILP,~\cite{Muggleton-Raedt:JLP94,Cropper-Dumancic:JAIR22}), which focuses on
learning \emph{logic programs} inductively. Concretely, the state-of-the-art ILP system \popper~\cite{cropper2021learning} shares a similar workflow with \tool: in a ``generate-test-prune'' loop, both systems use ASP to generate the candidate and prune the search. There are two main differences between \tool and \popper: 
(1) the preferred candidate (among all correct solutions) in ILP is the most general program because both positive and negative examples are given; while in \tool, the output is the conjunction of most specific formulas that satisfies all positive examples, because ``logical true'' is always a valid general formula in absence of negative examples. 
(2) the properties of FO formulas expose more domain knowledge to be encoded by ASP for pruning compared to Horn clauses.
\vspace{-5pt}
\paragraph{Other Approaches for Invariant Inference.} 
Among the techniques shown in \autoref{sec:overview:tool},
techniques from three tools (in the \textbf{Other techniques} category) have not been encoded
into \tool yet.
Two of them share similarities in how they improve the inference
process: \folsep~\cite{Koenig-al:PLDI20} is an inference framework
building on a first-order logic \emph{separability} solver;
\ictpo~\cite{Goel-al:NFM21} is an invariant inference tool based on the
relation between the \emph{symmetry} of satisfied FO structures and the
necessary \emph{quantification} in satisfying formulas in FOL.
Both of them are based on \pdr~\cite{Jhala-Schmidt:VMCAI11}, where
the generalisation of FO structures is performed based on a small set
of examples (usually less than 10, in contrast with \duoai, \swiss,
and \tool, where the generalisation is among thousands of examples).
From a technical perspective, their techniques can be encoded
into \tool, since both are SAT-based. However, the scalability
of their techniques make them not generally applicable to large sets of examples, which we will discuss
later.
%


%
%
\scimitar~\cite{Schultz-al:CoRR}, unlike most of the related works
where a ``global'' inductive invariant is inferred, is a tool that
infers ``local'' invariants for each transition state in the
distributed system protocol.
It builds an \emph{inductive proof graph} that abstracts the protocols,
and locally synthesises the invariants for each node in the graph,
where the local synthesis problem is an instance of our FO synthesis.
%

\vspace{-5pt}
\paragraph{Discussion on Inductive Generalisation.}

The least general generalisation~\cite{plotkin1970note} is known to be
a foundational result in the field of bottom-up inductive logic
programming, where the given programs (bottoms) are generalised to the
general program (top). In contrast, \tool is a top-down approach by
``generate-and-test''.
Notably, in the domain of distributed systems invariant inference, 
the authors of the \swiss tool~\cite{Hance-al:NSDI21}
described their ``failed attempt'' to use constraint solving for
bottom-up generalisation (without a detailed explanation).
This is coincident with our observation that the bottom-up generalisation (like \folsep and \ictpo) is not scalable to large sets of examples compared to the
top-down approach.
We provide our own explanation: given a whole search space,
any example can be regarded as a constraint to prune the search; while
the cost for the constraints can be considered linear in the number of examples, the
benefit of pruning decreases as the number of constraints increases. 
This brings a possible further work to combine the top-down and bottom-up
approaches by using subsets of examples for utilising the bottom-up generalisation.

Outside the domain discussed above, \cite{Krogmeier-Madhusudan:POPL22} is a notable work from automata theory that studies the learnability (\wrt generalisation) of formulas in finite variable logic. It does not focus on the concrete algorithm in practice, but show potential to extend \tool to logics more general than FOL.

\vspace{-10pt}

\section{Conclusion}
\label{sec:conclusion}

\vspace{-5pt}

In this work, we proposed a unified ASP-based framework \tool for synthesising formulas in first-order logic from examples.
%
%
To do so, we used ASP as a framework for implementing inductive formula synthesis offering constraint solving to encode the search and rule-based knowledge to prune the search space.
Using our ASP-based encoding, we proposed \emph{orthogonal slices}---a novel technique that significantly accelerates formula synthesis.
%
%
Finally, we have shown that declaratively capturing the essence of different approaches for formula synthesis in ASP enables a more efficient and composable solution.


\bibliographystyle{eptcs}
\bibliography{refs}

\appendix


\section{An Example of \autoref{def:fosynthesis} in \autoref{sec:overview:problem}}
\label{app:example}

This section provides an example of the FO synthesis problem.
\begin{example}[A tiny example]
    Considering a FO signature \(\Sigma = \langle C, R, F, S \rangle\) with $C$ and $F$ empty, $R = \{p, q, r\}$, and $S = \{X\}$. The synthesis problem is defined with the inputs:
  \begin{itemize}
      \item the set of FO formulas is restricted to the form of
        \(\forall X: lit_1\) or \(\forall X: lit_1 \lor lit_2\), which is a disjunction of one or two literals with one universal quantifier. Without the pruning, $\Omega_0$ contains $(2*3)+(2*3)^2 = 42$ formulas in total.
        \item the set of FO structures is $\sigma = \{M_1, M_2\}$, where $M_1$ and $M_2$ are two models over $\Sigma$ sharing the universe $\{x_0, x_1, x_2\} \in X $ and the interpretations of $p$, $q$, and $r$ are: 
        \begin{enumerate}
            \item $M_1$: [$p^{M_1} = \{x_0,~x_1\}$, $q^{M_1} = \{x_1,~x_2\}$, $r^{M_1} = \emptyset$]
            \item $M_2$: [$p^{M_2} = \{x_0,~x_1\}$, $q^{M_2} = \{x_2\}$, $r^{M_2} = \{x_1\}$]    
        \end{enumerate}
  \end{itemize}
  And the output of the synthesis problem based on \autoref{def:fosynthesis} is a set of formulas \[\Phi = \{\forall X. p(X) \lor q(X), \forall X. p(X) \lor\neg r(X) \}.\]
\end{example}

As a simple case of the sliced-template, our algorithm will first find satisfied formula in \(\forall X: lit_1\) (and fails), then checking the satisfied formula in \(\forall X: lit_1 \lor lit_2\) (with the two formulas successfully found). The inputs of \tool, the bounded FO search space and the real traces of distributed system protocols, are much more complex. An interested reader can refer to \texttt{configs/} and \texttt{traces/} folders in \url{https://github.com/verse-lab/FORCE} for the real instances.

\section{Additional Statistics of \autoref{sec:eval:combination}}
\label{app:statistics}

\begin{figure}[t]
  \centering

  \begin{subfigure}[t]{0.45\textwidth}
      \centering
      \begin{tikzpicture}[scale=1]
          \begin{axis}[
              title={Flexible Paxos},
              xlabel={Runtime (s)},
              ylabel={\scriptsize Size of output},
              legend pos=south west,
              legend style={font=\scriptsize},
              grid=major,
              xmin=0, xmax=300,
              ymin=0, ymax=900,
              width=\textwidth,
              height=0.7\textwidth,
          ]
          \addplot[
              only marks,
              mark=*,
              mark size = 1pt,
              color=blue,
          ] coordinates {
              (223.6,816) (250.5,816) (245.6,816) (176.2,816) (212,816)
          };
          \addlegendentry{\flyvy}
          \addplot[
              only marks,
              mark=*,
              mark size = 1pt,
              color=red,
          ] coordinates {
              (252.2,800) (270.3,796) (255.8,707) (188,717) (255.5,755)
          };
          \addlegendentry{\flyvy+~\duoai}
          \end{axis}
      \end{tikzpicture}
  \end{subfigure}
  \begin{subfigure}[t]{0.45\textwidth}
      \centering
      \begin{tikzpicture}[scale=1]
          \begin{axis}[
              title={Paxos},
              xlabel={Runtime (s)},
              ylabel={\scriptsize Size of output},
              legend pos=south west,
              legend style={font=\scriptsize},
              grid=major,
              xmin=0, xmax=1000,
              ymin=0, ymax=1300,
              width=\textwidth,
              height=0.7\textwidth,
          ]
          \addplot[
              only marks,
              mark=*,
              mark size = 1pt,
              color=blue,
          ] coordinates {
              (630,1260) (652.1,1260) (989.7,1260) (992.9,1260) (892.8,1260)
          };
          \addlegendentry{\flyvy}
          \addplot[
              only marks,
              mark=*,
              mark size = 1pt,
              color=red,
          ] coordinates {
              (751.7, 1106) (540.9, 1103) (671.9, 1080) (922, 1092) (780.7, 1092)
          };
          \addlegendentry{\flyvy+~\duoai}
          \end{axis}
      \end{tikzpicture}

  \end{subfigure}
  \caption{Optimising \flyvy via \duoai's output.}
  \label{fig:benchmark2}
\end{figure}

As said in \autoref{sec:eval:combination},
there are only two non-trivial protocols in \flyvy's benchmark that
have quantifier alternation, for which we can produce comparable
results. For both protocols, we execute the original \flyvy and the
optimised \flyvy for 5 times with  the sizes of the
output (been discussed) and the runtime shown in \autoref{fig:benchmark2}. 
Performance-wise, we did not achieve notable improvement on the
runtime of \flyvy, which is not unexpected, since the bottleneck of
\flyvy is not the synthesis; instead, the SMT solving in \flyvy
dominates the runtime and makes the runtime unstable.

\end{document}
